\documentclass[twocolumn,prl,aps,superscriptaddress,showpacs,amsmath,amssymb]{revtex4}
\usepackage{amsfonts}
\usepackage{bbm}

\usepackage{graphicx}
\usepackage{dcolumn}
\usepackage{bm}

\begin{document}

\title{Sequence-dependent spin-selective tunneling along double-stranded DNA}
\author{Ai-Min Guo}
\affiliation{Institute of Physics, Chinese Academy of Sciences, Beijing 100190, China}
\author{Qing-feng Sun}
\email{sunqf@iphy.ac.cn}
\affiliation{Institute of Physics, Chinese Academy of Sciences, Beijing 100190, China}

\begin{abstract}

We report spin-selective tunneling of electrons along natural and artificial double-stranded DNA (dsDNA) sandwiched by nonmagnetic leads. The results reveal that the spin polarization strongly depends on the dsDNA sequence and is dominated by its end segment. Both genomic and artificial dsDNA could be efficient spin filters. The spin-filtering effects are sensitive to point mutation which occurs in the end segment. These results are in good agreement with recent experiments and are robust against various types of disorder, and could help for designing DNA-based spintronic devices.

\end{abstract}


\maketitle

The charge transport along DNA molecule has received significant attention from scientific researchers over the past two decades.\cite{dc,erg,gjc1} In addition to electric charges, the DNA molecule could be also used to manipulate the electron spin. It was reported that self-assembled monolayers of double-stranded DNA (dsDNA) can discriminate the spin of photoelectrons.\cite{rsg,gb} These electrons transmitted through the dsDNA monolayers are highly polarized at room temperature and the spin-filtering effects are enhanced with increasing the DNA length.\cite{gb} Moreover, it was demonstrated that even single dsDNA could be efficient spin filter.\cite{xz} The underlying physical mechanism arises from the combination of the dephasing, the SO coupling, and the chirality of the DNA molecule.\cite{gam1} However, the spin polarization vanishes if the dsDNA was changed into single-stranded DNA or damaged by ultraviolet light.\cite{gb,xz,gam1}

The nitrogenous bases guanine (G), adenine (A), cytosine (C), and thymine (T), which are four basic ingredients of the DNA molecule, can constitute thousands of various sequences. While natural DNA molecule can be extracted from the cells of all living organisms, the artificial one could be synthesized in any desired sequence. It was shown that the DNA molecule with different sequences could present any transport behavior of conducting, semiconducting, and insulating.\cite{zyp,se,gx,sjd} One may thus expect that different dsDNA would display diverse spin-filtering effects. Indeed, the study of spin transport along various dsDNA will provide valuable information to the physical mechanism and the biological processes, and opens up its potential applications in molecular spintronics. In this Letter, we explore spin-selective tunneling of electrons through the dsDNA connected by normal-metal leads. Based on a model Hamiltonian which includes the SO coupling and the dephasing, the conductance and the spin polarization are calculated for a variety of dsDNA. Here, the DNA molecules involve genomic and artificial ones as well as those employed in the experiments.\cite{gb,xz} The sequences of several typical DNA samples are listed in Table~\ref{tab:table1}. The genomic dsDNA is extracted from the sequence of human chromosome 22 (chr22),\cite{note1} while the artificial dsDNA is taken as random sequence and substitutional one, e.g., Nickel mean (nm), Copper mean (cm), and Triadic Cantor (tc).\cite{me} All of the substitutional DNA sequences are constructed by initiating from one seed and following a substitution rule. For instance, the nm1 sequence is formed by adopting base A as the seed and the substitution rule A$\rightarrow$AGGG, G$\rightarrow$A.

\begin{table*}
\caption {The sequences of the DNA molecules. Here, only the sequence along one strand is presented, while the other can be derived according to Watson-Crick base-pairing rules: G pairs with C, and A with T. The first three terms are the DNA molecules adopted in the experiments, rd1, rd2, and rd3 are the random sequences, hc1, hc2, and hc3 are the chr22-based sequences, and the last four terms are the substitutional ones.} \label{tab:table1}
\begin{tabular}{ll}
\hline
Name  & DNA sequence\\
\hline
sq-26 & TTTGTTTGTTTGTTTGTTTTTTTTTT \\
sq-40 & TCTCAAGAATCGGCATTAGCTCAACTGTCAACTCCTCTTT \\
sq-50 & TACTCTACCTTCTCAAGAATCGGCATTAGCTCAACTGTCAACTCCTCTTT \\
rd1   & CAATGCAGTCTATCCACCTGACGGACCCCGACCCGCCTTT \\
rd2   & CAATGCAGTCTATCCACCTGACGGACCCCGACCCGGCTTT \\
rd3   & CAATGCAGTCTATCCACCTGACGGACCCCGACCCGCCATT \\
hc1   & TAAATAAATAAATAAATAAATAAAATAAATAAAAGCCTTT \\
hc2   & GGGCCCTGAGGCATGGGCCCAGAAGCATTCCTGTCCCCTT \\
hc3   & AGCTGGGGAGCAGGGCTCCACTCTGGGAGGGGGGCAGCCT \\
nm1   & AGGGAAAAGGGAGGGAGGGAGGGAAAAGGGAAAAGGGAAA \\
nm2   & ATTTAAAATTTATTTATTTATTTAAAATTTAAAATTTAAA \\
cm    & GAAGGGAAGAAGAAGGGAAGGGAAGGGAAGAAGAAGGGAA \\
tc    & GAGAAAGAGAAAAAAAAAGAGAAAGAGAAAAAAAAAAAAA \\
\hline
\end{tabular}
\end{table*}

From the study of numerous dsDNA, we find that the spin filtration efficiency presents strong dependence on the DNA sequence and is mainly determined by the end segment with several base-pairs. Both chr22-based and random dsDNA could be very efficient spin filters, while the substitutional one exhibits large spin polarization and conductance. Besides, the spin-filtering effects are sensitive to point mutation which takes place in the end segment of the dsDNA. The high spin polarization still holds even under the environment-induced on-site energy disorder and twist angle disorder. These results could be beneficial for building up DNA-based spintronic devices.

The spin transport along the dsDNA can be described by the Hamiltonian: ${\cal H}={\cal H}_{\rm DNA} + {\cal H}_{\rm so} +{\cal H}_d  +{\cal H}_{\rm lead}+{\cal H}_{c},$\cite{gam1,gam2} where ${\cal H} _{\rm DNA}= \sum_{j=1}^2(\sum_{n=1}^N  \varepsilon_ {jn} c_{jn}^\dag c_{jn}+ \sum_{n=1}^{N-1} t_{jn}c_{jn}^\dag c_{jn+1}) + \sum_{n=1}^N \lambda_n c_{1n}^\dag c_{2n} +\mathrm{H.c.}$ is the Hamiltonian of two-leg ladder model, with $n$ the base-pair index, $j$ the strand label, and $N$ the DNA length. $c_{jn} ^\dag= (c_{jn\uparrow}^\dag, c_{ jn \downarrow } ^\dag)$ is the creation operator, $\varepsilon_ {jn} $ is the on-site energy, $t_{jn}$ is the intrachain hopping integral, and $\lambda_n $ is the interchain hybridization interaction. The second term ${\cal H}_{\rm so}=\sum_{j n}\{ i t_{\rm so} c_{jn}^\dag \sigma _n ^ {(j)} c_{jn+1}+ \mathrm{H.c.} \}$ is the SO Hamiltonian, which stems from the double helix distribution of the electrostatic potential of the dsDNA.\cite{gam1} $t_{\rm so}$ is the SO coupling strength and $\sigma_n ^{(j ) } =[\sigma_ x ( \sin \varphi_{jn} +\sin \varphi_{jn+1})- \sigma_ y (\cos \varphi_{jn}+\cos \varphi_ {jn+1})] \sin \theta_{jn} +2\sigma_z \cos\theta_{jn}$, with $\sigma_ {x,y,z} $ the Pauli matrices, $\varphi _{jn}$ the cylindrical coordinate of the base, and $\theta _{jn}$ the helix angle between base $n$ and $n+1$ in the $j$th strand. In equilibrium position of the dsDNA, $\varphi _{jn}= (n-1) \Delta\varphi$ and $\theta_ {jn} = \theta$ with $\Delta\varphi$ the twist angle. The third one ${\cal H}_d= \sum_{jnk} ( \varepsilon_ {jnk} b_{jnk}^\dag b_{jnk}  + t_d b_{jnk}^\dag c_{jn} +\mathrm {H.c.} )$ is the Hamiltonian of the B\"{u}ttiker's virtual leads and their coupling with each base of the dsDNA, simulating the phase-breaking processes due to the inelastic scattering with phonons and counterions.\cite{bya2,bya3} The last two terms ${\cal H}_{\rm lead} +{\cal H}_c= \sum_{k,\beta=L,R} \varepsilon_{\beta k} a_{\beta k}^\dag a_{\beta k} +  \sum_{jk} ( t_L a_{L k}^\dag c_{j 1} +t_R a_{R k} ^ \dag c_{j N} + \mathrm {H.c.} )$ represent the real leads, and the coupling between these leads and the dsDNA, respectively. Finally, the conductances for spin-up $(G_ \uparrow)$ and spin-down $(G_ \downarrow)$ electrons can be calculated by using the Landauer-B\"{u}ttiker formula.\cite{gam1} The spin polarization is $P_s=(G_ \uparrow- G_ \downarrow )/ (G_\uparrow+ G_\downarrow)$. Since the current is flowing from the left real lead to the right one, the terminal of the dsDNA attached to the former is named the beginning, while the other terminal is called the end.

For the dsDNA, $\varepsilon_{jn}$ is chosen as the ionization potential with $\varepsilon_ {\rm G}= 8.3$, $\varepsilon_ {\rm A}= 8.5$, $\varepsilon_ {\rm C}= 8.9$, and $\varepsilon_ {\rm T}= 9.0$, $t_ {jn}$ between identical neighboring bases is taken as $t_{\rm GG}=0.11$, $t_{\rm AA}=0.22$, $t_{\rm CC}=-0.05$, and $t_{\rm TT}=-0.14$, and $\lambda_n =-0.3$. These parameters are extracted from the experimental results\cite{hns,dd,lj1,tab} and first-principles calculations\cite{cem,gfc3,vaa,zh,sk,hlgd,av} with the unit eV. $t_ {jn}$ between different neighboring bases X and Y is set to $t_{\rm XY}=(t_{\rm XX}+ t_{\rm YY})/ 2$, in accordance with first-principles results.\cite{vaa,zh,sk,hlgd} The helix angle and the twist one are $\theta=0.66$ rad and $\Delta \varphi ={\frac \pi 5}$. The SO coupling is estimated to $t_{\rm so}=0.01$. For the real leads, the parameters $\Gamma_L = \Gamma_R =1$ are fixed, while for the virtual ones, the dephasing strength is $\Gamma_ d=0.006$.

\begin{figure}
\includegraphics[width=0.41\textwidth]{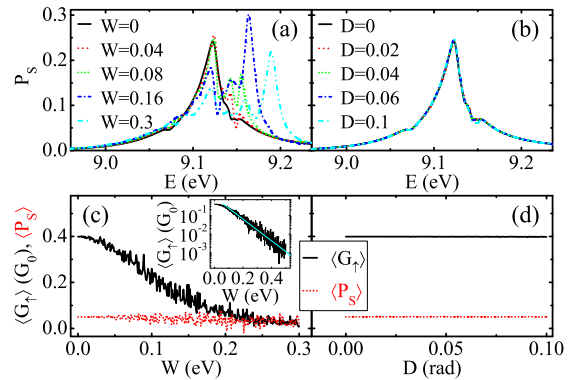}
\caption{\label{fig:one} Energy-dependent $P_s$ for poly(A)-poly(T) under the on-site energy disorder with degree $W$ (a) and of the twist angle disorder with degree $D$ (b). $\langle G_\uparrow \rangle$ and $\langle P_s\rangle$ vs $W$ (c) and vs $D$ (d). The inset of (c) shows $\langle G_\uparrow \rangle$ in a wider range of $W$ and the dependence can be fitted by the function $\langle G_\uparrow \rangle \propto 10^{-\alpha W}$ with $\alpha =5.80\pm0.09$ (cyan line). $\langle G_\uparrow \rangle$ and $\langle P_s\rangle$ are averaged in the energy region $[9.04, 9.32]$. All of the results are performed for single disorder configuration with $N=40$. Here, $G_0 =e^2/h$ is the quantum conductance.}
\end{figure}

It was reported that the ionization potential of the base is affected significantly by both counterions\cite{brn,zy} and hydration.\cite{ksk,yx,bl} Consequently, the environmental effects can be properly considered by varying the on-site energies. A random variable $w_{jn}$ is added in each $\varepsilon_{jn}$ to simulate the stochastic population of these counterions and water molecules around the dsDNA, with $w_{jn}$ uniformly distributed within the range $[-\frac W 2, \frac W 2]$ and $W$ the disorder degree. Fig.~\ref{fig:one}(a) shows the spin polarization $P_s$ of poly(A)-poly(T) under the on-site energy disorder, as a function of the energy $E$. It clearly appears that $P_s$ is large for homogeneous poly(A)-poly(T) and is sufficiently robust against the on-site energy disorder. This can be further demonstrated in Fig.~\ref{fig:one}(c), where the averaged spin polarization $\langle P_s\rangle$ is shown. One notices that $\langle P_s\rangle$ fluctuates around its equilibrium value of $5.0\%$ at $W=0$ and the oscillation amplitude is enhanced by $W$. Furthermore, a new energy region of high $P_s$ becomes more distinct in the case of larger $W$ [see the curves of $W=0.16$ and $0.3$ in Fig.~\ref{fig:one}(a)]. On the other hand, the averaged conductance $\langle G_ \uparrow\rangle$ is decreased by increasing $W$ as expected, due to the disorder-induced Anderson localization effect. The curve of $\langle G_ \uparrow \rangle $-$W$ can be fitted well by a simple function $\langle G_\uparrow \rangle \propto 10^{-\alpha W}$ [see inset of Fig.~\ref{fig:one}(c)].

Besides the on-site energy disorder, each base will waver around its equilibrium position at finite temperature. In this situation, it is reasonable to plus a random variable $d_{jn}$ in each $\varphi_{ jn}$, with $d _{jn}$ distributed in the region $[-\frac D 2, \frac D 2 ]$ and $D$ the disorder degree. By considering constant radius $R$ of the dsDNA and arc length $l_a$ between successive bases to account for the rigid sugar-phosphate backbone,\cite{gam1,gj} the helix angle $\theta _{jn}$ will be modulated according to $l_a\cos \theta_{jn} =R (\varphi_{jn+1}- \varphi_{jn})$ and the fluctuations are disregarded in the intrachain hopping integral as a first approximation.\cite{sk,gr,rs1} It can be seen from Fig.~\ref{fig:one}(b) that the curves of $P_s$-$E$ are superposed with each other in the context of the twist angle disorder only. Accordingly, no fluctuations could be observed in the curve of $\langle P_s\rangle$-$D$ [Fig.~\ref{fig:one}(d)]. Besides, $\langle G_ \uparrow \rangle$ will not be changed with $D$, because the SO coupling is much smaller than the hopping integral. Therefore, poly(A)-poly(T) remains an efficient spin filter even under the on-site energy disorder and the twist angle disorder.

\begin{figure}
\includegraphics[width=0.36\textwidth]{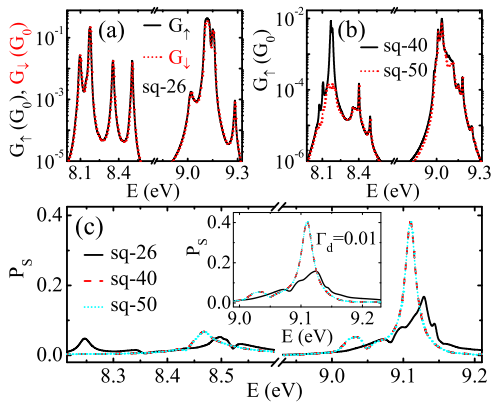}
\caption{\label{fig:two} Energy-dependence of $G_\uparrow $, $G_ \downarrow $, and $P_s$ for the dsDNA used in the experiments. (a) $G_{\uparrow}$ and $G_{\downarrow}$ for the sq-26 sequence. (b) $G_{\uparrow}$ for the sq-40 and sq-50 sequences. (c) $P_s$ for all three dsDNA. The inset of (c) displays $P_s$ with $\Gamma_d =0.01$.}
\end{figure}

Then we investigate the spin transport through aperiodic dsDNA in the absence of external environment-induced disorder. Our results still hold if this disorder is included. Let us first discuss the spin transport properties of the dsDNA used in the experiments.\cite{gb,xz} Fig.~\ref{fig:two}(a) displays the conductances of spin-up ($G_ \uparrow$) and spin-down ($G_ \downarrow$) electrons for sq-26 sequence, while Fig.~\ref{fig:two}(b) plots $G_ \uparrow$ for sq-40 and sq-50 sequences. As compared with homogeneous dsDNA,\cite{gam1} the energy spectrum of aperiodic dsDNA is also separated into the highest occupied molecular orbital (HOMO) and the lowest unoccupied molecular orbital (LUMO). The conductance is declined by increasing the DNA length, because the electrons experience stronger scattering in longer dsDNA.

In addition, one can see from Fig.~\ref{fig:two}(a) that the discrepancy between $G_ \uparrow$ and $G_ \downarrow$ is more distinct for the LUMO band than the HOMO one. Thus, $P_s$ is larger in the former band than the latter one [Fig.~\ref{fig:two}(c)]. Moreover, $P_s$ at $E=9.11$ is, respectively, $9.6\%$, $38\%$, and $38\%$ for the sq-26, sq-40, and sq-50 sequences, in good agreement with the experiment.\cite{gb} In fact, the obtained $P_s$ is also consistent with the experimental results by adopting different $\Gamma_ d$ from the region $[0.003, 0.01]$, e.g., see $P_s$ of $\Gamma_ d=0.01$ in the inset of Fig.~\ref{fig:two}(c). Besides, although the conductances between the sq-40 and sq-50 sequences are very different, their spin polarizations are almost identical and the difference between the two $P_s$ is within $10^{-6}$ range, due to the fact that the sq-40 sequence is the end part of the sq-50 sequence. This suggests that the spin filtration efficiency of the dsDNA is mainly controlled by its end segment, which will be substantiated below.

\begin{figure}
\includegraphics[width=0.39\textwidth]{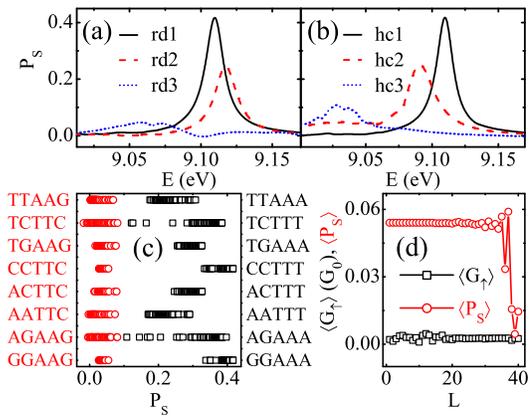}
\caption{\label{fig:three} Energy-dependent $P_s$ for the random dsDNA (a) and for the chr22-based one (b). (c) Distribution of $P_s$ at $E=9.11$ for various random dsDNA with large $P_s$ (right part) and with small $P_s$ (left part) as a comparison. The results are extracted from $10^5$ DNA samples. Here, only the end segment in the first strand is shown (two sides) and can be obtained for the second one according to the base-pairing rules. (d) $\langle G_ \uparrow \rangle$ and $\langle P_s \rangle$ vs mutation position $L$ for the rd1 sequence.}
\end{figure}

Next we turn to study the spin polarization of the random and chr22-based dsDNA. Figs.~\ref{fig:three}(a) and \ref{fig:three}(b) plot $P_s$ vs $E$ for several typical random and chr22-based sequences, respectively. It is clear from the curves of rd1 and hc1 that both random and chr22-based sequences could be very efficient spin filters with $P_s$ achieving $40\%$. From a statistical study of numerous dsDNA with extremely high $P_s$, it reveals that these sequences are terminated by the segment ``CCTTT/GGAAA'' in their ends [Fig. \ref{fig:three}(c)]. We emphasize that all of the investigated dsDNA with $N=40$ will exhibit very high $P_s$ around $40 \%$ if their end segments are replaced by ``CCTTT/GGAAA''. Besides, the dsDNA could be also very efficient spin filter if it has other end segments, as shown in Fig. \ref{fig:three}(c), where a distribution of $P_s$ at $E=9.11$ is displayed for different random dsDNA with 16 end segments. It clearly appears that $P_s$ is always large for these dsDNA [see the right part in Fig. \ref{fig:three}(c)], although $P_s$ will vary in a finite range. The dsDNA remains efficient spin filter if it is ended by the moiety ``(C)$_{m}$TT/(G)$_{m}$AA'' with $m$ the integer (see the curve of hc2). However, $P_s$ can be dramatically reduced by altering the end segment, even if its last base-pair is changed [see the left part in Fig. \ref{fig:three}(c)]. These are due to the fact that the charge will gradually lose its phase and spin memory while transmitting along the dsDNA. The longer distance the charge propagates, the larger the loss of its memory. Accordingly, the spin filtration efficiency of the dsDNA is dominated by its end segment containing several base-pairs.

\begin{figure}
\includegraphics[width=0.28\textwidth]{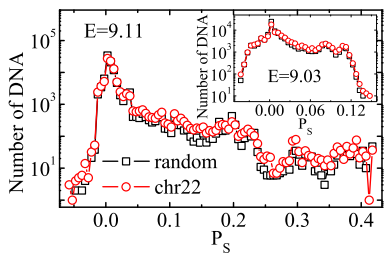}
\caption{\label{fig:four} Distribution functions of $P_s$ for the random and chr22-based dsDNA at $E=9.11$. The inset shows the corresponding statistics of $P_s$ at $E=9.03$. Here, $N=40$.}
\end{figure}

To further verify the aforementioned point, we introduce point mutation in the dsDNA, where only one base-pair is modified and replaced by another.\cite{sct} Here, the point mutation is defined by switching the complementary bases within single base-pair.\cite{note2} We focus on $P_s$ of the random dsDNA in Fig.~\ref{fig:three}(a). The rd2 and rd3 sequences are derived by introducing the point mutation in the rd1 one.\cite{note2} One notes that $P_s$ is reduced more significantly if the mutation position is closer to the last base-pair of the rd1 sequence. The largest $P_s$ is decreased from $42\%$ for the rd1 sequence to $25\%$ and $4.7\%$ for the rd2 and rd3 sequences, respectively. $\langle P_s\rangle$ and $\langle G_\uparrow \rangle$ are shown as a function of the mutation position $L$ in Fig.~\ref{fig:three}(d), where the average is obtained within the energy region $[8.98,9.18]$. It is clear that $\langle P_s\rangle$ does not change if the mutation occurs in the very beginning of the rd1 sequence, and fluctuates more strongly if the mutation position becomes closer to its end. $P_s$ is very small if the point mutation takes place within the last three base-pairs, due to the identical sign between $t_{1n}$ and $t_{2n}$.\cite{gam1} In contrast, $\langle G_\uparrow \rangle$ fluctuates more obviously if the mutation occurs in the beginning of the sequence. And the fluctuation amplitude is more severe in the curve of $\langle P_s\rangle$-$L$ than $\langle G_\uparrow \rangle$-$L$, indicating that the spin polarization is much more sensitive to the modification of the base-pair in the dsDNA than the conductance. In this perspective, the spin transport along the dsDNA may be related to mutation detection in the biological processes and could be beneficial for DNA sequencing.\cite{zm}

Figure~\ref{fig:four} shows the statistical properties of $P_s$ at fixed energy for the random and chr22-based dsDNA with $10^5$ samples. It clearly appears that $P_s$ can vary from $42\%$ to negative, implying that the spin polarization direction of the charges transmitted through the dsDNA could be reversed by modifying its sequence. And one can see that many DNA molecules exhibit high $P_s$. From a statistical perspective, the chr22-based dsDNA has more efficient spin filters than the random one. For instance, the number of the dsDNA, of which $P_s$ is larger than $30\%$ (20\%), is 458 (1436) and 667 (2020) for the random dsDNA and the chr22-based one, respectively. This can be further demonstrated in the inset of Fig.~\ref{fig:four}, where one notices that the curve of the chr22-based dsDNA is always higher than that of the random one for $P_s>1.3\%$. However, there are also many dsDNA with small $P_s$ at fixed energy. This is attributed to the fact that: (1) $P_s$ depends on $E$ that the energy region of large $P_s$ may differ from one sample to another [Figs.~\ref{fig:three}(a) and \ref{fig:three} (b)]; (2) the electrons may be not polarized exactly along the helix axis for each dsDNA and the actual spin polarization could be larger.

\begin{figure}
\includegraphics[width=0.39\textwidth]{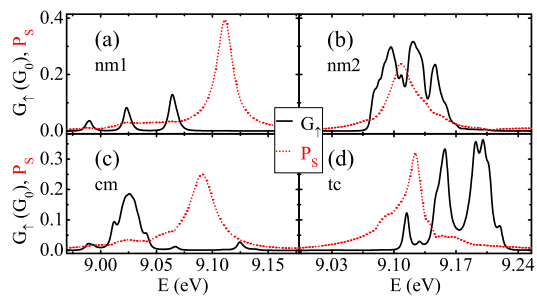}
\caption{\label{fig:five} Energy-dependence of $G_ \uparrow$ and $P_s$ for several substitutional sequences of DNA molecules.}
\end{figure}

Finally, we study the spin polarization of the substitutional sequences of DNA molecules, of which the electronic properties have been investigated previously.\cite{gam3,rs2} Fig.~\ref{fig:five} shows $P_s$ and $G_ \uparrow$ for several substitutional dsDNA. It is clear that both $P_s$ and $G_ \uparrow$ are very large for these dsDNA. Therefore, besides homogeneous DNA molecules, other aperiodic DNA sequences can be also efficient spin filters with high spin polarization and conductance.

In summary, we investigate the quantum spin transport through different dsDNA contacted by nonmagnetic leads. We find that the spin polarization strongly depends on the dsDNA sequence and is mainly determined by the end segment. Both natural and artificial dsDNA could be very efficient spin filters. Our results could motivate further experimental studies on DNA spintronics.

This work was financially supported by NBRP of China (2012CB921303 and 2009CB929100) and NSF-China under Grants Nos. 10974236 and 11074174.


\begin{references}
\bibitem{note1}  We consider the largest segment in human chromosome 22 sequence, which is retrieved from the National Center for Biotechnology Information (accession number: NT011520).
\bibitem{note2} The rd2 and rd3 sequences are transformed from the rd1 one. We obtain the former two sequences by substituting the 36th base-pair C/G with G/C and the 38th base-pair T/A with A/T, respectively.
\end{references}
\end{document}